\documentclass[showkeys,showpacs]{revtex4}
\usepackage{times}
\usepackage{epsfig,amssymb,amsfonts,amsmath}
\usepackage[usenames]{color}

\usepackage{amssymb}
\usepackage{amsmath}
\usepackage{amsfonts}
\usepackage{graphicx}
\usepackage{ulem} 
\usepackage{cancel}
\usepackage{extarrows}
\providecommand{\av}[1]{\left\langle #1 \right\rangle} %
\providecommand{\sqbr}[1]{\left[ #1 \right]} %

\begin{document}

\title[ ]{Comment on ``Troublesome aspects of the Renyi-MaxEnt treatment'' by A. Plastino, M.C. Rocca and F. Pennini}
\author{$^{1}$Thomas Oikonomou}
\email{thomas.oikonomou@nu.edu.kz}
\author{$^{2}$G. Baris Bagci}
\email{gbb0002@hotmail.com}

\affiliation{$^{1}$Department of Physics, School of Science and Technology, Nazarbayev University, 53 Kabanbay Batyr Ave., Astana 010000, Kazakhstan}
\affiliation{$^{2}$Department of Materials Science and Nanotechnology Engineering, TOBB University of Economics and Technology, 06560 Ankara, Turkey}
\keywords{Entropy maximization, Functional analysis, Tsallis/R\'enyi entropy}
\pacs{05.20.-y, 05.70.-a, 89.70.Cf}

\begin{abstract}
Plastino, Rocca and Pennini [Phys. Rev. E \textbf{94} (2016) 012145] recently stated that the R\'enyi entropy is not     suitable for thermodynamics by using functional calculus, since it leads to anomalous results unlike the Tsallis entropy. We first show that the Tsallis entropy also leads to such anomalous behaviours if one adopts the same functional calculus approach. Second, we note that one of the Lagrange multipliers is set in an \textit{ad-hoc} manner in the functional calculus approach of Plastino, Rocca and Pennini. Finally, the explanation for these anomalous behaviours is provided by observing that the generalized distributions obtained by Plastino, Rocca and Pennini does not yield the ordinary canonical partition function in the appropriate limit and therefore cannot be considered as genuine generalized distributions.

\end{abstract}

\eid{ }
\date{\today }
\startpage{1}
\endpage{1}
\maketitle


In Ref. \cite{PRP}, the authors obtain the relation $S_\text{R} = \ln Z$ (see Eq. (4.15) therein) by which they conclude that the R\'enyi entropy $S_\text{R}$ is not suitable for thermodynamics, since it has only dependence on the partition function $Z$ without an additional dependence on the average internal energy $\left\langle U \right\rangle $. According to Ref. \cite{PRP}, the Tsallis entropy $S_q$ is immune of such a defect. However, it is easy to find a similar relation also for the Tsallis entropy. Consider Eq. (2.18) in Ref. \cite{PRP}, multiply it by the probability distribution $P$ and then integrate to obtain
\begin{eqnarray}\label{T1}
\lambda_{2} = \frac{q \int_{M} P^{q} d \mu}{q-1}-\lambda_{1}\left\langle U \right\rangle
\end{eqnarray}
which shows that the Lagrange multipliers $\lambda_{1}$ and $\lambda_{2}$ are coupled also for the Tsallis case where $q$ is the Tsallis parameter as usual. Then substituting Eq. (\ref{T1}) back into Eq. (2.18) in Ref. \cite{PRP} and setting $\lambda_{1} = - \beta q $, we obtain exactly the same distribution in Eq. (4.7) in Ref. \cite{PRP} but now $\beta$ in Eq. (4.7) reads $\beta_{q}=\frac{\beta}{\int_{M} P^{q} d \mu}$ for the Tsallis case. Proceeding further in a similar manner to Ref. \cite{PRP}, we obtain the relation below
\begin{eqnarray}\label{T2}
S_q = \ln_q (Z_{q}) \,,
\end{eqnarray}
where $\ln_q (x)$ is the $q$-deformed logarithm \cite{Oiko1} and $Z_{q}$ is the concomitant expression for the Tsallis distribution.
One may think that $\lambda_2$ in Eq. (\ref{T1}) is a functional of $P$. Although it seems so \textit{prima facie}, it is simply not. Adopting Eq. (\ref{T1}) and proceeding exactly as in section IV of Ref. \cite{PRP}, after some algebra one obtains $\int_{M} P^{q} d \mu = Z_{q}^{1-q}$, with $Z_q=\int_{M}\sqbr{1+\beta\,(Z_q)^{q-1}\,(U-\av{U})}^{\frac{1}{1-q}}\mathrm{d}\mu$ so that Eq. (\ref{T1}) now becomes
\begin{eqnarray}\label{eq3}
\lambda_{2}= \frac{q Z_{q}^{1-q}}{q-1} - \lambda_{1} \left\langle U \right\rangle\,.
\end{eqnarray}
Therefore, there is no actual dependence of $\lambda_2$ on $P$. Note that there is no obstacle preventing the Lagrange multipliers to depend on the partition function itself as can be seen from Eqs. (2.20) or (2.21) in Ref. \cite{PRP}. Finally, note that Eq. (\ref{T2}) is possible also due to the fact that we have $\int_{M} P^{q} d \mu = Z_{q}^{1-q}$.

Eq. (\ref{T2}) shows that the essential link between statistical mechanics and thermodynamics is lost also in the case of the Tsallis entropy, since both Tsallis entropy and R\'enyi entropy yield similar equations ($S_\text{R} = \ln Z$ and $S_q = \ln_q (Z_{q})$, respectively) without the explicit appearance of the average internal energy as it should \cite{PRP}. As a side remark, note that the equation $S_q = \ln_q (Z_{q})$ is exactly the one obtained from the escort averaging procedure of the Tsallis entropy through ordinary calculus (see Eq. (27) in Ref. \cite{Mendes}). In this regard, the viewpoint presented in Ref. \cite{PRP} implicitly supports the claim that the escort averaging should be avoided, since it severs the link between statistical mechanics and thermodynamics.

The second closely related issue is the determination of $\lambda_{1}$ through functional calculus. The inspection of Eq. (1) in this comment (and Eq. (4.4) in Ref. \cite{PRP}) indicates that the Lagrange multipliers $\lambda_{1}$ and $\lambda_{2}$ are coupled after the maximization procedure. However, these equations do not uniquely yield $\lambda_{1}$. Once $\lambda_{1}$ is chosen in an \textit{ad-hoc} manner, $\lambda_{2}$ can be determined via Eq. (1) in this comment or Eq. (4.4) in Ref. \cite{PRP}. Therefore, the question remains as to how the expression $\lambda_{1} = - \beta \alpha$ (see Eq. (4.6) in Ref. \cite{PRP}) is uniquely obtained in Ref. \cite{PRP}.

The third issue is again related to the relation $S_\text{R} = \ln Z$ presented as Eq. (4.15) in Ref. \cite{PRP}. The authors indicate that the R\'enyi entropy $S_R$ does not reduce to the Boltzmann-Gibbs (BG) entropy for $\alpha \rightarrow 1$ at the end of Section IV in Ref. \cite{PRP}. However, this is utterly impossible, since they adopt the usual R\'enyi entropy in Eq. (2.1) in Ref. \cite{PRP} and it is well-known that the R\'enyi entropy $S_\text{R}$ does reduce to the BG entropy for $\alpha \rightarrow 1$.

As a final remark, we consider again the relation $S_\text{R} = \ln Z$ but now in the limit $\alpha \rightarrow 1$. In the aforementioned limit, $S_R$ becomes BG entropy so that the relation $S_\text{R} = \ln Z$ becomes
\begin{eqnarray}\label{T3}
S_\text{BG} = -\int_{M} P_{1} \ln (P_{1}) d \mu= \ln Z_{1} \,,
\end{eqnarray}
where $P_{1}$ and $Z_{1}$ denote the probability distribution and the partition function in Eqs. (4.7) and (4.8) in Ref. \cite{PRP} in the $\alpha \rightarrow 1$ limit. The inspection of the former two equations indicates that in this limit we have the ordinary canonical distribution as it should i.e.
\begin{eqnarray}\label{T4}
P_{1} = \frac{e^{-\beta U}}{\int_{M} e^{- \beta U} d \mu } \,.
\end{eqnarray}
The substitution of Eq. (5) into Eq. (4) yields
\begin{eqnarray}\label{T5}
Z_{1} = e^{ \beta \left\langle U\right\rangle }{\int_{M} e^{- \beta U} d \mu }
\end{eqnarray}
instead of the well-known canonical distribution $Z_\text{BG}= \int_{M} e^{- \beta U} d \mu$. Note that Eq. (\ref{T5}) can also be directly obtained from Eq. (4.8) in Ref. \cite{PRP} by taking the limit $\alpha \rightarrow 1$ \cite{note1}. 
In fact, if one substitutes $Z_1$ in Eq. (\ref{T5}) back into the right hand side of Eq. (\ref{T3}), we see that the relation between statistical mechanics and thermodynamics is restored i.e. one obtains $S_\text{BG} = \ln Z_\text{BG}+\beta \left\langle U\right\rangle $ (see Eq. (3.13) in Ref. \cite{PRP}) as it should. Note that if $Z_1$ in Eq. (\ref{T3}) would indeed be the ordinary canonical partition function, then, one should have deduced that the statistical mechanics implies $\beta \left\langle U\right\rangle = 0$ for all classical physical systems independent of their Hamiltonian as a result of the relation in Ref. \cite{PRP} i.e. $S_\text{R} = \ln Z$.

It is important to understand the implication of Eq. (\ref{T5}) above: It simply implies that the authors in Ref. \cite{PRP} used a generalized distribution which yields the ordinary canonical partition function to be $e^{ \beta \left\langle U\right\rangle }{\int_{M} e^{- \beta U} d \mu } $ contrary to the well-known textbook result $\int_{M} e^{- \beta U} d \mu  $. However, a generalization should at least preserve the ingredients of the former theory it is built on in the appropriate limit.
Hence, both Tsallis and Renyi entropies are not problematic for the use of thermodynamics if the appropriate generalized distribution is considered.

%



\begin{thebibliography}{20}
\bibitem{note1} Note that the probability distribution remains the same as the canonical one in the $\alpha \rightarrow 1$ limit in Ref. \cite{PRP} (consider using Eqs. (4.7) and (4.8) together in the aforementioned limit in Ref. \cite{PRP}), since the term $e^{ \beta \left\langle U\right\rangle }$, appearing both in the nominator and denominator, cancels out.
\end{thebibliography}
\end{document}